\documentclass[twocolumn,prc,superscriptaddress,unsortedaddress,showpacs,preprintnumbers,amsmath,amssymb]{revtex4}
\usepackage{graphicx}
\usepackage{dcolumn}
\usepackage{bm}
\usepackage{CJK}
\usepackage{CJKulem}
\usepackage{txfonts}
\usepackage[dvipdfmx,bookmarks=true,colorlinks,%
citecolor=blue,linkcolor=blue,anchorcolor=blue,filecolor=blue,urlcolor=blue,%
           ]{hyperref}

\def\beq{\begin{equation}}
\def\eeq{\end{equation}}
\def\bea{\begin{eqnarray}}
\def\eea{\end{eqnarray}}

\def\fun#1#2{\lower3.6pt\vbox{\baselineskip0pt\lineskip.9pt
  \ialign{$\mathsurround=0pt#1\hfil##\hfil$\crcr#2\crcr\sim\crcr}}}

\usepackage[scaled=0.92]{helvet}
\usepackage[T1]{fontenc}
\usepackage{textcomp}

\begin{document}
\begin{CJK*} {GBK} {song}
\preprint{}

\title{Generalized isobaric multiplet mass equation and
its application to the Nolen-Schiffer anomaly}

\author{J. M. Dong}\affiliation{Institute of Modern Physics, Chinese
Academy of Sciences, Lanzhou 730000, China}
\author{Y. H. Zhang}
\affiliation{Institute of Modern Physics, Chinese Academy of
Sciences, Lanzhou 730000, China}
\author{W. Zuo}
\affiliation{Institute of Modern Physics, Chinese Academy of
Sciences, Lanzhou 730000, China} \affiliation{School of Physics,
University of Chinese Academy of Sciences, Beijing 100049, China}
\author{J. Z. Gu}
\affiliation{China Institute of Atomic Energy, P. O. Box 275(10),
Beijing 102413, China}
\author{L. J. Wang}
\affiliation{School of Physics and Astronomy, Shanghai Jiao Tong
University, Shanghai 200240, China}
\author{Y. Sun}\email[ ]{sunyang@sjtu.edu.cn }
\affiliation{Institute of Modern Physics, Chinese Academy of
Sciences, Lanzhou 730000, China} \affiliation{School of Physics and
Astronomy, Shanghai Jiao Tong University, Shanghai 200240, China}
\affiliation{Collaborative Innovation Center of IFSA, Shanghai Jiao
Tong University, Shanghai 200240, China}

\date{\today}

\begin{abstract}

The Wigner Isobaric Multiplet Mass Equation (IMME) is the most
fundamental prediction in nuclear physics with the concept of
isospin. However, it was deduced based on the Wigner-Eckart theorem
with the assumption that all charge-violating interactions can be
written as tensors of rank two. In the present work, the
charge-symmetry breaking (CSB) and charge-independent breaking (CIB)
components of the nucleon-nucleon force, which contribute to the
effective interaction in nuclear medium, are established in the
framework of Brueckner theory with AV18 and AV14 bare interactions.
Because such charge-violating components can no longer be expressed
as an irreducible tensor due to density dependence, its matrix
element cannot be analytically reduced by the Wigner-Eckart theorem.
With an alternative approach, we derive a generalized IMME (GIMME)
that modifies the coefficients of the original IMME. As the first
application of GIMME, we study the long-standing question for the
origin of the Nolen-Schiffer anomaly found in the Coulomb
displacement energy of mirror nuclei. We find that the
naturally-emerged CSB term in GIMME is largely responsible for
explaining the Nolen-Schiffer anomaly.

\end{abstract}

\maketitle

{\it Introduction.} The similarity of proton and neutron masses and
approximate symmetry of nucleon-nucleon interactions under the
exchange of the two kinds of nucleons lead to the concept of isospin
\cite{Heisenberg32,Wigner37}. At the isospin-symmetry limit, the
charge-symmetry requires that the free proton-proton interaction
$v_{pp}$ excluding the Coulomb force is equal to the neutron-neutron
$v_{nn}$, while the charge-independence requires that the
neutron-proton interaction $v_{np} = (v_{nn}+v_{pp})/2$~\cite{Cou2}.
However, the nucleon-nucleon scattering data suggested that $v_{nn}$
is slightly more attractive than $v_{pp}$, and $v_{np}$ is stronger
than $(v_{nn}+v_{pp})/2$~\cite{np1,Machleidt01}. In real nuclear
systems where many-body effects are important \cite{Warner06},
isospin symmetry breaking has long been an active research theme
connected to different subfields, for examples, in understanding the
precise values of the Cabbibo-Kobayashi-Maskawa (CKM) mixing matrix
elements between the $u$ and $d$ quarks \cite{Hardy0509,Towner08},
the changes in nuclear structure near the $N=Z$ line due to
charge-violating nuclear force \cite{INC2,INC3,INC4,INC-exp}, and
the influence in nova nucleosynthesis \cite{MSU16}.

Isobaric nuclei with the same mass number $A$, total isospin $T$,
and spin-parity $J^{\pi}$, but different $T_z=(N-Z)/2$, form an
isobaric multiplet. The Wigner isobaric multiplet mass equation
(IMME)~\cite{IMME1}
\begin{equation}
\text{ME}(A,T,T_{z})=a+bT_{z}+cT_{z}^{2}  \label{AA}
\end{equation}
provides a relationship for mass excesses of an isobaric multiplet,
where $a$, $b$ and $c$ are the coefficients depending on $T$ and
reduced matrix elements. This quadratic form of IMME turns out to
work remarkably well for almost all isobaric multiplets where data
exist~\cite{Review1,Review2,Review3}. Hence it becomes a powerful
tool to predict unknown masses, particularly those of very
neutron-deficient nuclei important for the astrophysical
$rp$-process~\cite{rp}. Modern radioactive beam facilities can
provide the testing grounds of the validity of the
IMME~\cite{Mass53,Mass20}, from which one may learn about the
effective forces for nuclear many-body systems
\cite{force1,force2,SHN1,SHN2,astrophysics}.

The IMME is regarded to be valid for {\it any} charge-violating
interactions, with the Coulomb interaction to be the dominant
contributor. The values of $b$ and $c$ in Eq. (\ref{AA}), which are
determined experimentally, can potentially yield individual
information on violations of the charge symmetry and
charge-independence \cite{Cou2}. However, the proven validity of the
IMME does not in itself provide any direct information on the nature
of the charge-violating nuclear interaction. In shell-model
calculations, such interaction \cite{Ormand1989,Lam2013} are {\it
added} to an isospin-conserving Hamiltonian, with the
charge-symmetry breaking (CSB) or charge-independent breaking (CIB)
components in the strong nuclear force fitted to data. In this Rapid
Communication, we consider the contributions of CSB and CIB derived
from nuclear medium in the effective nucleon-nucleon interaction.
Due to density dependence of the charge-violating components,
additional terms emerge as compared to the Wigner original IMME,
leading to a generalized isobaric multiplet mass equation (GIMME).
As the first application of GIMME, the binding-energy difference
between two members of a multiplet, defined as the Coulomb
displacement energy (CDE), is examined. The long-standing problem of
the Nolen-Schiffer anomaly~\cite{Nolen} in CDE is addressed by using
our new formulae, without the need of involving any empirical terms.

{\it The effective CSB and CIB interactions in nuclear matter.} In
the study of nuclear matter with the assumption of isospin
conservation in nuclear forces, the energy per nucleon is generally
given as a function of density $\rho =\rho _{n}+\rho _{p}$ and
isospin asymmetry $\beta =(\rho _{n}-\rho _{p})/\rho $, via $ E(\rho
,\beta )= E(\rho ,0)+S_2(\rho )\beta ^{2}+\mathcal{O(}\beta
^{4})$~\cite{Review11,Review12,Review13,Review14}, where the
density-dependent $S_2(\rho )$ is the widely-studied 2nd-order
symmetry energy coefficient. If one does not neglect the CSB and CIB
components, additional terms appear
\begin{eqnarray}
E(\rho ,\beta ) &=&E(\rho ,0)+S_{0}^{\text{(CIB)}}(\rho )+S_{1}^{\text{(CSB)}%
}(\rho )\beta  \nonumber\\
&&+\left[ S_{2}(\rho )+S_{2}^{\text{(CIB)}}(\rho )\right] \beta ^{2}+%
\mathcal{O(}\beta ^{3}). \label{EOS}
\end{eqnarray}
Specifically, the effective CSB interaction, namely the CSB
component of the effective nucleon-nucleon interaction, gives rise
to the 1st-order symmetry energy coefficient, defined as $
S_{1}(\rho )=\partial E(\rho ,\beta )/{\partial \beta }|_{\beta
=0}$, while the CIB interaction solely contributes to even-order
ones. In other words, $S_{1}^{\text{(CSB)}}$
($S_{2}^{\text{(CIB)}}$) measures the CSB (CIB) effect in nuclear
medium.

\begin{figure}[htbp]
\begin{center}
\includegraphics[width=0.40\textwidth]{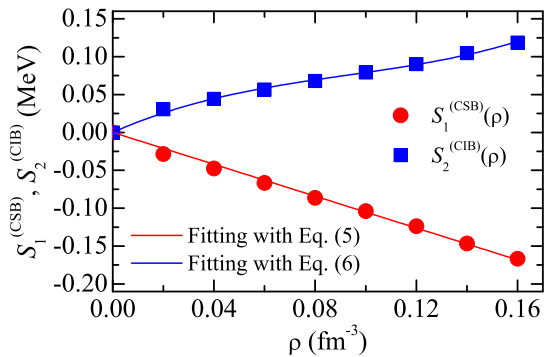}
\caption{(Color online) Density-dependent $S_1^{\text{(CSB)}}(\rho)$ (dots)
and $S_2^{\text{(CIB)}}(\rho)$ (squares) of nuclear matter obtained with the
Brueckner-Hartree-Fock approach adopting the AV18 along with AV14
bare interactions.
The curves represent the fittings with Eqs.
(\ref{coe1},\ref{coe}).}\label{fig1}
\end{center}
\end{figure}

Contributions of the CSB and CIB components in a bare potential to
the effective two-body interaction in nuclear matter can be obtained
by solving the Bethe-Goldstone equation in the Brueckner theory with
the AV18 interaction. AV18 contains explicit charge-dependence and
charge-asymmetry supplemented to the AV14 potential~\cite{AV18}. To
achieve a reliable accuracy, we determine the $S_{1}^{\text{(CSB)}}$
term with the formula
\begin{equation}
\frac{E(\rho ,\beta )-E(\rho ,-\beta )}{2\beta
}|_{\text{AV18}}=S_{1}^{\text{(CSB)}}(\rho ),
\end{equation}
in order to cancel out the systematical uncertainty effectively. In
addition, the $n$-$p$ mass difference in nucleonic kinetic energy
leads to a small part of 1st-order symmetry energy~\cite{Dong2016},
and is incorporated into the CSB effect in the present discussion.
Similarly, the even-order symmetry energy coefficients originating
from the CIB interaction are extracted by adopting both the AV18 and
AV14 potentials via
\begin{equation}
\frac{E(\rho ,\beta )+E(\rho ,-\beta )}{2}|_{\text{AV18}}-E(\rho ,\beta )|_{%
\text{AV14}}=S_{0}^{\text{(CIB)}}(\rho )+S_{2}^{\text{(CIB)}}(\rho
)\beta ^{2}.
\end{equation}
$S_{0}^{\text{(CIB)}}$ is an additional energy induced by the CIB
interactions in symmetric nuclear matter, referred to as the
zeroth-order symmetry energy coefficient. As a constant for an
isobaric multiplet, $S_{0}^{\text{(CIB)}}$ can be absorbed into
$E(\rho ,0)$, playing no role in the present discussion.
Figure~\ref{fig1} illustrates the density-dependent
$S_{1}^{\text{(CSB)}}(\rho )$ and $S_{2}^{\text{(CIB)}}(\rho )$,
which are found to be much smaller than the widely-investigated
2nd-order one $S_{2}(\rho )$, and therefore, have been completely
neglected in the study of nuclear matter. For the discussion below,
we perform polynomial fittings for $S_{1}^{\text{(CSB)}}(\rho )$ and
$S_{2}^{\text{(CIB)}}(\rho )$ obtained from the Brueckner theory
\begin{eqnarray}
S_{1}^{\text{(CSB)}}(\rho ) &=&a_{0}\rho , \label{coe1} \\
S_{2}^{\text{(CIB)}}(\rho ) &=&a_{1}\rho +a_{2}\rho ^{2}+a_{3}\rho
^{3},\label{coe}
\end{eqnarray}
with the resulting coefficients listed in Table~\ref{coeff}.

\begin{table}[h]
\label{table1} \caption{\label{coeff} The fitted coefficients of Eqs.
(\ref{coe1},\ref{coe}).}
\begin{ruledtabular}
\begin{tabular}{cccc}
$a_0$ (MeV$\cdot $fm$^{-3}$)  & $a_1$ (MeV$\cdot $fm$^{-3}$)  &  $a_2$ (MeV$\cdot $fm$^{-6}$)  & $a_3$ (MeV$\cdot $fm$^{-9}$) \\
\hline
$-1.05132$ & 1.49199  & $-10.96773$  & 39.58976 \\
\end{tabular}
\end{ruledtabular}
\end{table}

{\it The CSB and CIB effects in finite nuclei.} With the above
results derived for nuclear matter, we now build a Skyrme energy
density functional for the effective CSB and CIB interactions.
Considering the above Eqs. (\ref{coe1},\ref{coe}) obtained, we
construct the effective two-body CSB and CIB interactions by
\begin{eqnarray}
v_{\text{CSB}} &=&-2a_{0}P_{ij}^{\sigma }(\tau_{3,i}+\tau_{3,j})\delta (%
\overrightarrow{r_{i}}-\overrightarrow{r_{j}}) \\
v_{\text{CIB}} &=&-4(a_{1} +a_{2}\rho +a_{3}\rho ^{2})P_{ij}^{\sigma
}\delta (\overrightarrow{r_{i}}-\overrightarrow{r_{j}}),
\end{eqnarray}
where the $P_{ij}^{\sigma }$ is the spin exchange operator and
$\tau_3$ is the third-component of the Pauli operator. The local
density $\rho$ is evaluated at $(\overrightarrow{r_{i}} +
\overrightarrow{r_{j}})/2$, with $\overrightarrow{r_{i}}$ and
$\overrightarrow{r_{j}}$ being, respectively, the spacial
coordinates of the $i$-th and $j$-th nucleons. Accordingly, the
expressions for the corresponding energy density are given as
\begin{eqnarray}
\mathcal{H}_{\text{CSB}} &=&a_{0}(\rho _{n}^{2}-\rho _{p}^{2}), \\
\mathcal{H}_{\text{CIB}} &=&(a_{1}+a_{2}\rho +a_{3}\rho ^{2})(2\rho
_{n}^{2}+2\rho _{p}^{2}-\rho ^{2}),
\end{eqnarray}
and hence the energy per nucleon $\mathcal{H}_{\text{CSB}}/\rho$
$\left(\mathcal{H}_{\text{CIB}}/\rho \right)$ is exactly the
symmetry energy term $S_{1}^{\text{(CSB)}}(\rho )\beta$
$\left(S_{2}^{\text{(CIB)}}(\rho )\beta ^{2}\right)$ in
Eq.~(\ref{EOS}). Note that the isospin exchange operator
$P_{12}^{q}=\delta _{q1,q2}$ is assumed since the charge-mixing is
quite weak. Therefore, the 1st- and 2nd-order symmetry energy
coefficients, $a^{\text{(CSB)}}_{\text{sym,1}}(A,T_z)$ and
$a^{\text{(CIB)}}_{\text{sym,2}}(A,T_z)$ for finite nuclei, can be
calculated as corresponding density functionals
\begin{eqnarray}
a_{\text{sym,1}}^{\text{(CSB)}}(A,T_{z})
&=&\frac{1}{IA}\int_{0}^{\infty
}4\pi r^{2}\rho (r)S_{1}^{\text{(CSB)}}(\rho )\beta (r)dr,  \label{GG1} \\
a_{\text{sym,2}}^{\text{(CIB)}}(A,T_{z})
&=&\frac{1}{I^{2}A}\int_{0}^{\infty }4\pi r^{2}\rho
(r)S_{2}^{\text{(CIB)}}(\rho )\beta ^2 (r)dr.  \label{GG2}
\end{eqnarray}
In the above equations, $I=(N-Z)/A=2T_z/A$ denotes isospin asymmetry
of a nucleus, and $\beta(r) =(\rho _{n}(r)-\rho _{p}(r))/\rho(r)$ is
the local isospin asymmetry, with $\rho _{p}(r)$ and $\rho_{n}(r)$
being the proton and neutron density distribution, respectively.

We comment on how excited states in a given multiplet are calculated
in our theory, although these states do not appear in the discussion
of the present work. For an isobaric analog state (IAS) with $N-1$
neutrons and $Z+1$ protons ($N>Z$) whose $T$ is greater than
$|T_z|$, its wave function can be obtained by~\cite{IAS1,IAS2}
$|\text{IAS}\rangle
=|T,T_{z}=T-1\rangle=\frac{1}{\sqrt{2T}}T_{-}|\text{0}\rangle$,
where $T_-$ is the isospin lowering operator and $|\text{0}\rangle$
is the ground state of the parent nucleus belonging to a multiplet
with $T = T_z$ ($N$ neutrons and $Z$ protons). Due to the above
isospin-symmetry conserving operation, it naturally leads to $(\rho
_{n}+\rho _{p})_{\text{IAS}}=(\rho _{n}+\rho_{p})_{\text{parent}}$.
However, $T_{-}|\text{0}\rangle/\sqrt{2T}$ describes the IAS with
the zeroth-order approximation only which conserves isospin. Because
of the core polarization induced by the charge-violating
interactions, corresponding corrections should be
introduced~\cite{IAS2}. Consequently, $(\rho_{n}-\rho_{p})
_{\text{IAS}}-(\rho_{n}-\rho_{p})
_{\text{parent}}=-\frac{1}{T}\rho_{n,\text{parent}} ^{\text{exc.}}$
is obtained, where $\rho _{n,\text{parent}}^{\text{exc.}}$ is the
density of the excess neutrons in the parent nucleus. Thus with the
obtained nucleonic density distributions, the symmetry energies of
the IAS can be also computed by the above density functionals.

Since $a^{\text{(CSB)}}_{\text{sym,1}}(A,T_z)$ and
$a^{\text{(CIB)}}_{\text{sym,2}}(A,T_z)$ are related solely to the
nuclear force, one should perform many-body calculations excluding
the Coulomb force, which leads to
$a^{\text{(CSB)}}_{\text{sym,1}}(A,T_{z})=a^{\text{(CSB)}}_{\text{sym,1}}(A,-T_{z})$
and
$a^{\text{(CIB)}}_{\text{sym,2}}(A,T_{z})=a^{\text{(CIB)}}_{\text{sym,2}}(A,-T_{z})$
for mirror nuclei within an isobaric multiplet. Furthermore,
considering the fact that the CSB and CIB effects are small, we
treat them as perturbations. Consequently, both
$a^{\text{(CSB)}}_{\text{sym,1}}(A,T_{z})$ and
$a^{\text{(CIB)}}_{\text{sym,2}}(A,T_{z})$ are completely isolated
from the rest of the energy, and thus can be reliably extracted.

\begin{table}[h]
\label{table2} \caption{\label{SE} The 1st-order symmetry energy
coefficient $a^{\text{(CSB)}}_{\text{sym,1}}(A,T_z)$ [keV] (the first
three columns) and 2nd-order one
$a^{\text{(CIB)}}_{\text{sym,2}}(A,T_z)$ [keV] (the last three columns)
for finite nuclei obtained by Eqs. (\ref{GG1},
\ref{GG2}).}
\begin{ruledtabular}
\begin{tabular}{ccccccc}
Nuclei  & SLy4  &  SLy5  & KDE & SLy4  &  SLy5  & KDE \\
\hline
$^{20}$O & -40.7  & -40.0  & -43.1  & 20.3 & 19.9 & 21.1 \\
$^{53}$Ni & -107.5 & -106.2  & -109.7   & 86.8 & 85.7 & 86.4 \\
$^{208}$Pb& -111.9  & -112.0   & -116.1   & 79.9 & 80.1 & 82.5\\
\end{tabular}
\end{ruledtabular}
\end{table}

We now briefly discuss the calculated
$a^{\text{(CSB)}}_{\text{sym,1}}(A,T_z)$ and
$a^{\text{(CIB)}}_{\text{sym,2}}(A,T_z)$ for finite nuclei. The
Skyrme-Hartree-Fock-BCS approach with three interactions studied in
our previous work~\cite{Dong15}, i.e., the SLy4, SLy5 and KDE
interactions~\cite{SLY}, are employed to calculate the quantities of
Eqs. (\ref{GG1}, \ref{GG2}), in which the empirical gaps from
Ref.~\cite{Gap11} are applied. Table~\ref{SE} lists the calculated
results, taking $^{20}$O (a member of $A=20$ quintet), $^{53}$Ni (a
member of $A=53$ quartet), and a heavy nucleus $^{208}$Pb as
examples. Both $a^{\text{(CSB)}}_{\text{sym,1}}(A,T_z)$ and
$a^{\text{(CIB)}}_{\text{sym,2}}(A,T_z)$ are found to be weakly
model-dependent because different interactions generate nearly
identical nucleonic density profiles. For the members of isobaric
multiplets, such as $^{53}$Ni, the values of the 1st-order symmetry
energy term $
E^{\text{(CSB)}}_{\text{sym,1}}(A,T_z)=a^{\text{(CSB)}}_{\text{sym,1}}(A,T_z)I
A$ are very small due to their low isospin asymmetries $I$ and the
undersized $S_{1}^{\text{(CSB)}}(\rho )$. On the other hand,
$E^{\text{(CSB)}}_{\text{sym,1}}(A,T_z)$ for $^{208}$Pb can be as
large as $-5$ MeV. Apparently, the 2nd-order ones,
$E^{\text{(CIB)}}_{\text{sym,2}}(A,T_z)=a^{\text{(CIB)}}_{\text{sym,2}}(A,T_z)I^2
A$, are smaller.

We thus conclude that the 1st-order symmetry energy term should not
always be neglected in the calculations for neutron-rich nuclei. We
note, for example, that nuclear masses can be presently predicted by
employing macroscopic-microscopic mass models~\cite{MASS1,MASS2}
with an accuracy of several hundred keV. Furthermore, CSB
interaction has been shown to play an important role in nuclear
structure~\cite{INC3,INC4}. Our obtained effective interactions
including the symmetry-breaking components could be employed to
explore the relevant problems such as the charge-exchange reactions,
Gamow-Teller transitions, and $\beta$-decays. Up to now, shell-model
calculations for these quantities can only be performed by
introducing phenomenological symmetry-breaking terms with the
strengths fitted to data \cite{Kaneko17}.

{\it A generalized IMME including effective CSB and CIB
interactions.} In his derivation of Eq. (\ref{AA}), Wigner assumed
$|\alpha T T_z\rangle$ to be the eigenstate of the
charge-independent Hamiltonian $H_0$, with $\alpha$ for all
additional quantum numbers to specify this state. All
charge-violating two-body interactions, including the Coulomb
interaction $H_{\text{C}}$ among protons and $H_{\text{CSB+CIB}}$ of
CSB and CIB interactions, are treated by the first-order
perturbation. The total negative binding energy is given by
\begin{equation}
-\text{BE}(\alpha TT_{z})=\langle \alpha TT_{z}|H_{0}+H_{\text{C}}+H_{\text{CSB+CIB%
}}|\alpha TT_{z}\rangle,
\end{equation}
where $H_{\text{C}}$ and $H_{\text{CSB+CIB}}$ are assumed to be
written as tensors of rank two. With help of the Wigner-Eckart
Theorem for irreducible tensor, the perturbing terms can be neatly
expressed as reduced matrix elements and the coefficients involving
only $T$ and $T_z$.

However, in nuclear medium, $H_{\text{CSB+CIB}}$ becomes {\it
density-dependent} effective interaction. As a
result, it can no longer be expressed as an irreducible tensor, and
the corresponding perturbation energy $\langle \alpha
TT_{z}|H_{\text{CSB+CIB}}|\alpha TT_{z}\rangle$ does not have
analytic forms as in the case of the Coulomb interaction. When the
effective CSB and CIB interactions are present, the perturbation
energy in the present work is expressed as the symmetry energy terms
\begin{equation}
\langle \alpha TT_{z}|H_{\text{CSB+CIB}}|\alpha TT_{z}\rangle =a_{\text{sym,1}%
}^{\text{(CSB)}}(A,T_{z})IA+a_{\text{sym,2}}^{\text{(CIB)}}(A,T_{z})I^{2}A,
\end{equation}
with the zeroth-order symmetry energy coefficient absorbed into $a$.
One thus ends up with a generalized IMME (GIMME) in the form of
\begin{eqnarray}
\text{ME}(A,T,T_{z}) &=&a+\left( b_c+\Delta _{\text{nH}}+2a_{\text{sym,1}}^{\text{(CSB)}%
}(A,T_{z})\right) T_{z}  \label{HH} \nonumber\\
&&+\left(
c_{c}+\frac{4}{A}a_{\text{sym,2}}^{\text{(CIB)}}(A,T_{z})\right)
T_{z}^{2},
\end{eqnarray}
with $\Delta _{\text{nH}}=0.782$ MeV being the neutron-hydrogen mass
difference. As a mass equation beyond the original IMME, the
contribution from the effective charge-violating nuclear
interactions is now completely separated from that of the Coulomb
force, while the $T_{z}$-independent $b_c$ and $c_c$ in
Eq.~(\ref{HH}) are induced solely by the Coulomb interaction. The coefficients of $T_z$
and $T_z^2$ are no longer constants for a given multiplet. The
$T_{z}$-dependence of the new
$a^{\text{(CSB)}}_{\text{sym,1}}(A,T_z)$ and
$a^{\text{(CIB)}}_{\text{sym,2}}(A,T_z)$ terms, originating from the
CSB and CIB components of nuclear medium, are an explicit indication
of the breakdown of the original IMME. We remark that this
$T_{z}$-dependence is quite weak, supporting the general validity of
the original IMME \cite{IMME1} that has been tested against many
experimental data. Yet, under certain circumstances, the quadratic
form of the IMME may break down, and the underlying mechanism will
be discussed in further detail in a forthcoming paper.

\begin{figure}[htbp]
\begin{center}
\includegraphics[width=0.45\textwidth]{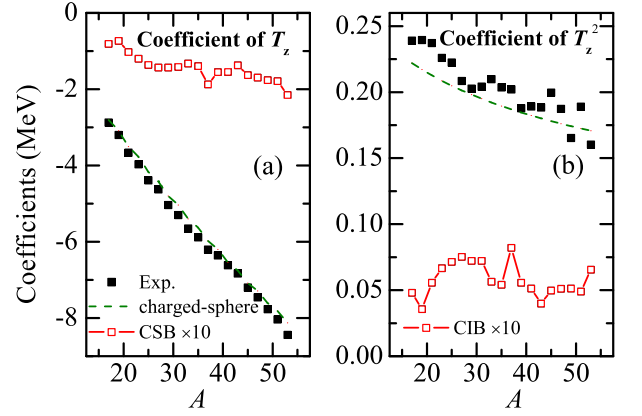}
\caption{Coefficients of (a) $T_z$ and (b)
$T^2_z$ in Eq. (\ref{AA}) extracted from the experimental data~\cite{Audi} for
the $T=3/2$ quartets. The calculated coefficients of the Coulomb
contribution (plus $\Delta _{\text{np}}$) from a simple
non-uniformly charged sphere with Eq. (\ref{Cou}) (dashed curves) are shown for
comparison. The contributions of the CSB
and CIB effects in Eq. (\ref{HH}), taking $T_z =T$ nuclei as
examples, are calculated by using
Eqs. (\ref{GG1}, \ref{GG2}) with the SLy4 interaction.}\label{fig2}
\end{center}
\end{figure}

We now discuss how much the corrections are actually introduced by
the CSB and CIB effects, and examine their systematic behavior. With
the assumption that the nucleus is treated as a non-uniformly
charged sphere~\cite{PD}, the Coulomb energy $E_c$ can be written as
\begin{equation}
E_{c}=\frac{3e^{2}}{5r_{0}A^{1/3}(1+\Delta )}\left[ Z(Z-1)-0.25\left(
1-(-1)^{Z}\right) \right] ,
\end{equation}
with $r_0=1.2$ fm, where the correction due to the last unpaired
proton~\cite{NP1960} is supplemented. The parameter $\Delta=5\pi^2
d^2/(6r_0^2A^{2/3})$ with $d \approx 0.55$ fm~\cite{PD} is
introduced to describe the effect of the surface diffuseness on the
Coulomb energy, which is a correction to the uniformly charge sphere
model~\cite{IAS1}, and the Coulomb interaction on the surface
asymmetry is ignorable for the $N\approx Z$ nuclei. Hence the
contributions of the Coulomb energy to the coefficients of $T_z$ and
$T_z^2$ are simply derived as
\begin{eqnarray}
b_{c} &=&\frac{3e^{2}}{5r_{0}A^{1/3}(1+\Delta )}\left[ (1-A)+\frac{%
(-1)^{A/2-T}-(-1)^{A/2+T}}{8T}\right] ,   \nonumber\\
\quad c_{c} &=&\frac{3e^{2}}{5r_{0}A^{1/3}(1+\Delta )}\left[ 1+\frac{%
(-1)^{A/2-T}+(-1)^{A/2+T}}{4(2T-1)}\right] . \label{Cou}
\end{eqnarray}

Figure~\ref{fig2} illustrates the coefficients of the $T_z$ and of
$T_z^2$ terms extracted from the masses of the $T=3/2$ isobaric
quartets~\cite{Audi}, and compares them with those given by a
non-uniformly charged sphere~\cite{PD}. The contributions of the CSB
and CIB effects in Eq. (15), taking $T_z= T$ nuclei as examples, are
also presented in Fig. 2 for comparison. The contribution of the CSB
effect to the coefficient of $T_z$ term increases roughly from $-80$
keV to $-220$ keV when $A$ goes up from 17 to 53, which is found to
be consistent with the estimations for the $T=1$ multiplets given in
Table 5.4  of Ref. \cite{CA}. In general, the CSB effect results in
a reduction of the coefficient of $T_z$ term by $2.0\%-3.1\%$, and
the CIB effect enhances the coefficient of $T_z^2$ term by
$1.6\%-4.4\%$. Note that, while the energy splitting among the
isobaric multiplet is predominately attributed to the Coulomb
interaction, clearly the corrections to the IMME have the CSB and
CIB origin.

{\it On the Nolen-Schiffer anomaly.} The Nolen-Schiffer anomaly
\cite{Nolen} (NSA) is a long-standing historical problem. The
Coulomb displacement energy (CDE) -- the difference in binding
energy between two members of a multiplet -- is directly related to
the IMME coefficients in Eq. (\ref{AA}): for adjacent members of a
multiplet one has $\text{CDE}(A,T,T_{z})=-b-c(2T_{z}+1)+\Delta
_{\text{nH}}$ \cite{Cou2}, where $T_z$ is taken for the isobar with
the larger proton number. It is an anomaly because when all the
corrections were taken into account, there remained a consistent
under-estimate of the CDE by about a few to ten percents
\cite{Nolen,IAS1,Auerbach1980}. Now with our GIMME, the CDE
expression is modified as
\begin{equation}
\text{CDE}(A,T,T_{z})=-b_{c}-c_{c}(2T_{z}+1)+\Delta _{\text{NSA}},
\end{equation}
where the new last term arising from the CSB and CIB components of
the nuclear medium is given by
\begin{eqnarray}
\Delta _{\text{NSA}} &=&-2a_{\text{sym,1}}^{\text{(CSB)}}(A,T_{z>})-\frac{%
4(2T_{z}+1)}{A}a_{\text{sym,2}}^{\text{(CIB)}}(A,T_{z>})  \notag \\
&&+2T_{z}\left[ a_{\text{sym,1}}^{\text{(CSB)}}(A,T_{z})-a_{\text{sym,1}}^{%
\text{(CSB)}}(A,T_{z>})\right]  \notag \\
&&+\frac{4T_{z}^{2}}{A}\left[ a_{\text{sym,2}}^{\text{(CIB)}}(A,T_{z})-a_{%
\text{sym,2}}^{\text{(CIB)}}(A,T_{z>})\right] ,  \notag \\
&\simeq &-2a_{\text{sym,1}}^{\text{(CSB)}}(A,T_{z>})-\frac{4(2T_{z}+1)}{A}a_{%
\text{sym,2}}^{\text{(CIB)}}(A,T_{z>}),
\end{eqnarray}
with $T_{z>}=T_z+1$. With $\Delta _{\text{NSA}}$, it becomes clear
that the CDE has contributions from CSB and CIB, in addition to the
Coulomb force. The CSB effect contributes
predominately in Eq. (19), whereas the CIB effect is much smaller,
particularly for heavier masses due to the $1/A$ dependence.
According to Fig.~\ref{fig2}, $\Delta _{\text{NSA}}$ accounts
for $2\%$-$3\%$ of the CDE for isobaric quartets, which, according to our calculation,
can add to CDE with 100-200 keV for $T_z=\pm1/2$ and 300-600 keV for
$T_z=\pm3/2$ mirror pairs. These amounts are qualitatively consistent with what
is needed to account for the Nolen-Schiffer anomaly, as discussed in Ref.~\cite{Cou2}.

\begin{table}[h]
\label{table3} \caption{\label{table1} The calculated $\Delta
_{\text{NSA}}=-2a^{\text{(CSB)}}_{\text{sym,1}}$ (in MeV)
due to the CSB effect for the $T=1/2$ mirror pairs in the $A=16$ and 40
regions, compared with other calculations for the study of the Nolen-Schiffer
anomaly.}\label{III}
\begin{ruledtabular}
\begin{tabular}{cccccc}
Nuclide  & SLy4  &  SLy5  & KDE & Ref.~\cite{SIII} & Ref.~\cite{Shlomo} \\
\hline
$^{15}$O-$^{15}$N & 0.16 & 0.16 & 0.16 &0.29& $0.16\pm0.04$   \\
$^{17}$F-$^{17}$O & 0.11  & 0.11  & 0.11 &0.11 & $0.31\pm0.04$ \\
$^{39}$Ca-$^{39}$K & 0.15 & 0.16 & 0.16 & 0.44 & $0.22\pm0.08$\\
$^{41}$Sc-$^{41}$Ca & 0.14 & 0.14 & 0.15 & 0.12&  $0.59\pm0.08$\\
\end{tabular}
\end{ruledtabular}
\end{table}

The CDE for a $T=1/2$ pair of mirror nuclei,
defined as $\text{CDE}(A,T=1/2)=\text{BE}(A,T_z=1/2)-\text{BE}(A,T_z=-1/2)$,
has been widely used to study the Nolen-Schiffer anomaly. In our method, the CDE is given by
$\text{CDE}(A)=-b_c-2a^{\text{(CSB)}}_{\text{sym,1}}$, and the CSB effect is simply
obtained with $\Delta
_{\text{NSA}}=-2a^{\text{(CSB)}}_{\text{sym,1}}$. Our calculated $\Delta
_{\text{NSA}}$ for nuclei near the closed shells with the magic numbers 8 and 20, compared with those based on the SIII Skyrme interaction~\cite{SIII} and a
calibrated independent-particle model~\cite{Shlomo} with
inclusion of many corrections for some extensively studied mirror
pairs, are listed in Table \ref{III}. It should be noted
that the results from the dominant Coulomb term and the
small corrections, such as the finite size of nucleons and
short-range correlation, exhibit considerable differences
between Refs.~\cite{SIII} and~\cite{Shlomo}, suggesting a
model-dependence character in the results. Moreover, the core-polarization correction, even
in its sign, presents a strong model-dependence~\cite{SIII,BARR}. Interestingly,
our results are found consistent with those in Ref.~\cite{SIII} (~\cite{Shlomo})
for particle (hole) nuclei. We emphasize, however, that our $\Delta
_{\text{NSA}}$ is completely separated from the Coulomb energy, and
in addition, our results are obtained without tuning any particular parameters.

\begin{figure}[htbp]
\begin{center}
\includegraphics[width=0.45\textwidth]{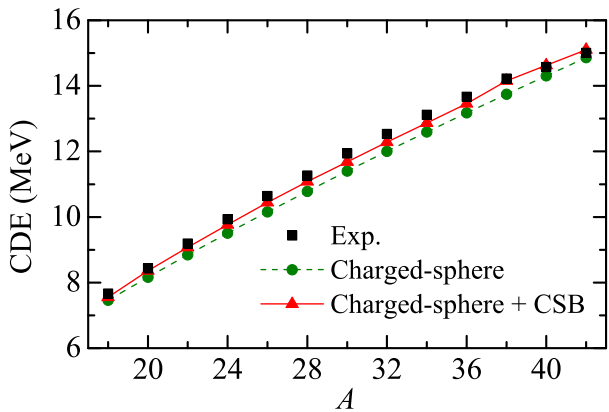}
\caption{(Color online) Comparison of calculated CDE of the $T=1$ mirror pairs
with experimental data~\cite{DATA2012}. The calculated results with and
without the second term in Eq. (\ref{CDET1}) are plotted as triangles and dots, respectively.
The first term in Eq. (\ref{CDET1}) is calculated using Eq. (\ref{Cou}), while the
second one using Eqs. (\ref{GG1}) with the SLy4 interaction.} \label{fig3}
\end{center}
\end{figure}

From our derivation, the CDE of a pair of mirror nuclei with $T=1$ becomes
\begin{equation}
\text{CDE}(A, T=1)=-2b_c-4a^{\text{(CSB)}}_{\text{sym,1}}(A,T_z=1).\label{CDET1}
\end{equation}
In the above expression, the contribution to the CSB effect is
directly obtained as $\Delta
_{\text{NSA}}=-4a^{\text{(CSB)}}_{\text{sym,1}}(A,T_z=1)$. In order
to compare our results with the CDE data, we compute the CDE with
the non-uniformly charged sphere model, with and without the second
term in Eq. (\ref{CDET1}). The results together with experimental data are presented in Fig.~\ref{fig3}. The difference between the two calculations is obvious. It can be seen that overall, the calculated CDE with inclusion of the CSB effect tends to describe the experimental data, where the CSB effect contributes an amount of $2\%-3\%$.

The origin of the Nolen-Schiffer anomaly has been studied by many
works (see, for example, Ref.~\cite{IAS1}) and is generally expected
to result mainly from the CSB effect. The (isospin)
symmetry-breaking terms are usually not included in normal
shell-model Hamiltonians, and therefore the answer to the anomaly
lies likely in a deeper level \cite{Schiffer}. In contrast to the
applied models based on the effects of the nucleon mass splitting or
meson mixing~\cite{MHS,Hatsuda1990,Cohen1991,Epele1992}, in our
framework, the CSB and CIB effects starting from nuclear medium are
established by employing the microscopic Brueckner theory without
any adjustable parameter.
Incidently, the triplet displacement energy (TDE) \cite{TDE} is
related to our new coefficients through
$\text{TDE}(A)=2c_c+8a_{\text{sym,2}}^{\text{(CIB)}}(T_z=1)/A$.
However, the $c_c$ coefficient cannot be well achieved with the
charged sphere model, as that in Fig.~\ref{fig2}. Different from CDE
of a mirror pair discussed above, the TDE originates from the
Coulomb force together with the CIB effect, where the latter
contributes about $3\%$. As the Nolen-Schiffer anomaly for CDE, the
Coulomb interaction alone cannot account for TDE, a conclusion
consistent with the shell model studies \cite{INC3,INC4}.

{\it Summary.} We have generalized the Wigner IMME by considering
the contributions of CSB and CIB derived in nuclear medium to the
effective nucleon-nucleon interaction, and used it to study the
Nolen-Schiffer anomaly. The main conclusions are as follows. i) The
density-dependent CSB and CIB interactions in nuclear matter,
characterized respectively by the symmetry energy coefficients
$S_{1}^{\text{(CSB)}}$ and $S_{2}^{\text{(CIB)}}$, were built within
the Brueckner theory with the bare interactions as inputs.
Therefore, our work bridges the charge-violating nuclear force in
free nucleons and that in nuclear medium. ii) With these results as
calibrations, we established the effective CSB and CIB interactions
in the Skyrme functions, and carried out the calculations of their
effects in finite nuclei. For neutron-rich nuclei, we found that the
1st-order symmetry energy term $E_{\text{sym,1}}(A)$ induced by the
CSB effect, which is generally dropped in nuclear mass calculations,
should not be neglected. iii) The perturbative Hamiltonian with the
density-dependent effective CSB and CIB interactions is no longer an
irreducible tensor, hence its matrix element cannot be analytically
reduced via the Wigner-Eckart theorem, as Wigner did \cite{IMME1}.
We derived the GIMME which presents new corrections to the original
Wigner IMME, where the contribution of the effective CSB and CIB
interactions is clearly separated from that of Coulomb force. iv) As
the first application of GIMME, the Nolen-Schiffer anomaly, which
has been a long-standing challenge to nuclear physics, was naturally
elucidated to a large extent to originate from the CSB effect, with
the needed correction of several hundreds keV being reproduced.

Finally, we note that our obtained CIB interaction in an effective Skyrme energy density functional describes only the ground-state properties for finite nuclei. The $J$-dependence of CIB (see, for example Refs. \cite{Kaneko17,Bentley15}) cannot be discussed here. This is however an important aspect of CIB, and should be investigated in the future within the present theory.

J. M. D. would like to thank Yu. A. Litvinov for helpful comments
and suggestions, and gratefully acknowledge the support of K. C.
Wong Education Foundation. Y. S. acknowledges the discussion with J.
Schiffer in his visit to the Argonne National Laboratory in July
2017. This work was supported by the National Natural Science
Foundation of China under Grants Nos. 11435014, 11775276, 11405223,
11675265, and 11575112, by the 973 Program of China under Grant No.
2013CB834401 and No. 2013CB834405, by the National Key Program for
S\&T Research and Development (No. 2016YFA0400501), by the Knowledge
Innovation Project (KJCX2-EW-N01) of Chinese Academy of Sciences, by
the Funds for Creative Research Groups of China under Grant No.
11321064, and by the Youth Innovation Promotion Association of
Chinese Academy of Sciences.

\end{CJK*}


\begin{thebibliography}{DONG}

\bibitem{Heisenberg32} W. Heisenberg, Z. Phys. {\bf 77}, 1 (1932).

\bibitem{Wigner37} E. Wigner, Phys. Rev. {\bf 51}, 106 (1937).

\bibitem{Cou2}
M. A. Bentley, S. M. Lenzi, Prog. Part. Nucl. Phys. {\bf 59}, 497
(2007).

\bibitem{np1}
E. M. Henley, in {\it Isospin in Nuclear Physics}, edited by D. H.
Wilkinson (North-Holland, Amsterdam, 1969).

\bibitem{Machleidt01} R. Machleidt, Phys. Rev. C {\bf 63}, 024001 (2001).

\bibitem{Warner06} D. D. Warner, M. A. Bentley, and P. Van Isacker, Nat. Phys. {\bf 2},
311 (2006).

\bibitem{Hardy0509} J. C. Hardy and I. S. Towner, Phys. Rev.
C {\bf 71}, 055501 (2005); {\bf 79}, 055502 (2009).

\bibitem{Towner08} I. S. Towner and J. C. Hardy, Phys. Rev. C {\bf 77}, 025501 (2008).

\bibitem{INC2}
A. P. Zuker, S. M. Lenzi , G. Mart\'inez-Pinedo, and A. Poves, Phys.
Rev. Lett. {\bf 89}, 142502 (2002).

\bibitem{INC3}
K. Kaneko, Y. Sun, T. Mizusaki, and S. Tazaki, Phys. Rev. Lett. {\bf
110}, 172505 (2013).

\bibitem{INC4}
K. Kaneko, Y. Sun, T. Mizusaki, and S. Tazaki, Phys. Rev. C {\bf
89}, 031302(R) (2014).

\bibitem{INC-exp} J. Henderson {\it et al.}, Phys. Rev. C {\bf 90},
051303(R) (2014).

\bibitem{MSU16}
M. B. Bennett {\it et al.}, Phys. Rev. Lett. {\bf 116}, 102502
(2016).

\bibitem{IMME1}
E. P. Wigner, {\it in Proc. of the R. A. Welch Foundation Conf. on
Chemical Research, Houston}, edited by W. O. Millikan (R. A. Welch
Foundation, Houston, 1957), Vol. 1; S. Weinberg and S. B. Treiman,
Phys. Rev. {\bf 116}, 465 (1959).

\bibitem{Review1}
W. Benenson and E. Kashy, Rev. Mod. Phys. {\bf 51}, 527 (1979).

\bibitem{Review2}
J. Britz, A. Pape and M. S. Antony, At. Data Nucl. Data Tables {\bf
69}, 125 (1998).

\bibitem{Review3}
Y. H. Lam {\it et al.}, At. Data Nucl. Data Tables {\bf 99}, 680
(2013);

\bibitem{rp}
A. Parikh, A. Parikh, J. Jose, C. Iliadis, F. Moreno, and T.
Rauscher, Phys. Rev. C {\bf 79}, 045802 (2009).

\bibitem{Mass53}
Y. H. Zhang {\it et al.}, Phys. Rev. Lett. {\bf 109}, 102501 (2012).

\bibitem{Mass20}
A. T. Gallant {\it et al.}, Phys. Rev. Lett. {\bf 113}, 082501
(2014).

\bibitem{force1}
F. Wienholtz {\it et al.}, Nature {\bf 498}, 346 (2013).

\bibitem{force2}
J. D. Holt, J. Men\'{e}ndez, and A. Schwenk, Phys. Rev. Lett. {\bf
110}, 022502 (2013).

\bibitem{SHN1}
E. Minaya Ramirez {\it et al.}, Science {\bf 337}, 1207 (2012)

\bibitem{SHN2}
D. Steppenbeck {\it et al.}, Nature {\bf 502}, 207 (2013).

\bibitem{astrophysics}
K. Blaum, Phys. Rep. {\bf 425}, 1 (2006).

\bibitem{Ormand1989}
W. E. Ormand and B. A. Brown, Nucl. Phys. A {\bf 491}, 1 (1989).

\bibitem{Lam2013}
Y. H. Lam, N. A. Smirnova, and E. Caurier, Phys. Rev. C {\bf 87},
054304 (2013).

\bibitem{Nolen}
J. A. Nolen, J. P. Schiffer, Annu. Rev. Nucl. Part. Sci. {\bf 19},
471 (1969).


\bibitem{Review11}
A. W. Steiner, M. Prakash, J. M. Lattimer, and P. J. Ellis, Phys.
Rep. {\bf 411}, 325 (2005).

\bibitem{Review12}
V. Baran, M. Colonna, V. Greco, and M. Di Toro, Phys. Rep. {\bf
410}, 335 (2005).

\bibitem{Review13}
B. A. Li, L. W. Chen, and C. M. Ko, Phys. Rep. {\bf 464}, 113
(2008).

\bibitem{Review14}
J. M. Lattimer and M. Prakash, Phys. Rep. {\bf 621}, 127 (2016).


\bibitem{AV18}
R. B. Wiringa, V. G. J. Stoks, and R. Schiavilla, Phys. Rev. C {\bf
51}, 38 (1995).

\bibitem{Dong2016}
J. M. Dong, W. Zuo and J. Z. Gu, Chin. Phys. Lett. {\bf 33}, 102101
(2016).

\bibitem{IAS1}
N. Auerbach, Phys. Rep. {\bf 98}, 273 (1983), and references cited
therein.

\bibitem{IAS2}
N. Auerbach, N. Van Giai, Phys. Rev. C {\bf 24}, 782 (1981).

\bibitem{Dong15}
J. Dong, W. Zuo, and J. Gu, Phys. Rev. C {\bf 91}, 034315 (2015).

\bibitem{SLY}
E. Chabanat, P. Bonche, P. Haensel, J. Meyer, R. Schaeffer, Nucl.
Phys. A {\bf 635}, (1998) 231; B. K. Agrawal, S. Shlomo, and V. K.
Au, Phys. Rev. C {\bf 72}, 014310 (2005).

\bibitem{Gap11}
P. M\"oller, J. R. Nix, W. D. Myers, and W. J. Swiatecki, At. Data
Nucl. Data Tables {\bf 59}, 185 (1995).

\bibitem{MASS1}
P. M\"oller, W. D. Myers, H. Sagawa, and S. Yoshida, Phys. Rev.
Lett. {\bf 108}, 052501 (2012).

\bibitem{MASS2}
N. Wang, M. Liu, X. Wu, J. Meng, Phys. Lett. B {\bf 734}, 215
(2014).

\bibitem{Kaneko17}
K. Kaneko, Y. Sun, T. Mizusaki, S. Tazaki, S. K. Ghorui, Phys. Lett.
B {\bf 773}, 521 (2017).


\bibitem{PD}
P. Danielewicz, Nucl. Phys. A {\bf 727}, 233 (2003).

\bibitem{NP1960}
S. Sengupta, Nucl. Phys. {\bf 21}, 542 (1960).

\bibitem{Audi}
M. MacCormick, G. Audi, Nucl. Phys. A {\bf 925}, 61 (2014).

\bibitem{CA}
N. Auerbach, J. Hufner, A. K. Kerman, and C. M. Shakin, Rev. Mod.
Phys. {\bf 44}, 48 (1972).

\bibitem{SIII}A. Poves, A. L. Cedillo and J. M. G. G\'omez, Nucl. Phys. A {\bf 293}, 397 (1977).

\bibitem{Shlomo}
S. Shlomo, Rep. Prog. Phys. {\bf 41}, 957 (1978).

\bibitem{BARR}
A. Barroso, Nucl. Phys. A {\bf 281}, 267 (1977).

\bibitem{Schiffer}
J. Schiffer, private communications.

\bibitem{Auerbach1980}
N. Auerbach, V. Bernard, and N. Van Giai, Phys. Rev. C {\bf 21}, 744
(1980); Nucl. Phys. A {\bf 337}, 143 (1980).

\bibitem{Hatsuda1990}
T. Hatsuda, H. Hogaasen, and M. Prakash, Phys. Rev. C {\bf 42}, 2212
(1990).

\bibitem{Cohen1991}
T. D. Cohen, R. J. Furnstahl, and M. K. Banerjee, Phys. Rev. C {\bf
43}, 357 (1991).

\bibitem{Epele1992}
L. N. Epele, H. Fanchlottl, C. A. Garc\'ia Canal, G. A. Gonz\'alez
Sprlnberg, Phys. Lett. B {\bf 277}, 33 (1992).

\bibitem{MHS}
M. H. Shahnas, Phys. Rev. C {\bf 50}, 2346 (1994).

\bibitem{DATA2012}
M. Wang {\it et al.}, Chin. Phys. C {\bf 36}, 1603 (2012).

\bibitem{TDE}
P. E. Garrett {\it et al.}, Phys. Rev. Lett. {\bf 87}, 132502
(2001).

\bibitem{Bentley15}
M. A. Bentley, S. M. Lenzi, S. A. Simpson, and C. Aa. Diget, Phys. Rev.
C {\bf 92}, 024310 (2015).

\end{thebibliography}
\end{document}